\documentclass[a4paper]{jpconf}
\usepackage{graphicx}

\newcommand{ \be }{\begin{equation}}
\newcommand{ \ee }{\end{equation}}
\newcommand{ \bea }{\begin{eqnarray}}
\newcommand{ \eea }{\end{eqnarray}}
\newcommand{ \la }{\langle}
\newcommand{ \ra }{\rangle}

\newcommand{\mean}[1]{\left\langle #1 \right\rangle} 
\newcommand{\meani}[1]{\overline{ #1 } }

\newcommand{ \dpti }{\Delta p_{t,1}}
\newcommand{ \dptj }{\Delta p_{t,2}}
\newcommand{ \dptdpt }{\dpti \dptj }
\newcommand{ \mdptdpt }{\mean{\dpti \dptj} }
\newcommand{ \midptdpt }{\meani{\dpti \dptj} }
\newcommand{ \bp }{{\bf p}}
\newcommand{ \DY }{{\Delta X}}
\newcommand{ \tR }{\tilde{R}}
\newcommand{ \mpt }{{\la p_t \ra}}

\usepackage{color}

\begin{document}
\title{Heavy ion collisions:
Correlations and Fluctuations in particle production.}

\author{Sergei A.~Voloshin        }
\address{Department of Physics and Astronomy, 
        Wayne State University, \\ 
        666 W. Hancock, Detroit, Michigan 48201}

\begin{abstract}
Correlations and fluctuations (the latter are directly related to 
the 2-particle correlations) is one of the important directions in 
analysis of heavy ion collisions. 
At the current stage of RHIC exploration, when the details matter, 
basically any physics question is  addressed  with help
of correlation techniques. 
In this talk I start with a general introduction to the correlation 
and fluctuation formalism and discuss weak and strong sides 
of different type of observables.
In more detail, I discuss the two-particle $p_t$ correlations/$\mpt$
fluctuations. In spite of not observing any dramatic
changes in the event-by-event fluctuations with energy, which would indicate a
possible phase transition, such correlations measurements remain an
interesting and important subject, bringing valuable information. 
Lastly, I show how radial flow can generate characteristic
azimuthal, transverse momentum and rapidity correlations, which could
qualitatively explain many of recently observed phenomena in nuclear
collisions.
\end{abstract}

\section{Introduction}

The physics of high energy heavy ion collisions attracts strong
attention of the physics community as creation of a new type of
matter, the Quark-Gluon Plasma, is expected in such collisions. 
During just a few years of the BNL RHIC operation many new phenomena 
has been observed,
such as strong elliptic flow~\cite{star-flow1}, and  
suppression of the high transverse momentum two particle back-to-back 
correlations~\cite{star-highpt}.
Note that both of the above mentioned discoveries at RHIC have
been made with help of correlations techniques: in this case analyzing
azimuthal correlations. 
Further understanding of the physics processes responsible for both
phenomena, such as what happens with the energy of the away-side 
jet or whether the observed constituent quark 
scaling~\cite{voloshinQM2002,star-q-scaling,molnar-voloshin} in elliptic flow
reflects the hadronization via quark coalescence, require further dedicated 
correlation studies. 
In general, as we try to understand the details of system evolution 
we will rely more and more on correlation measurements. 

Many  collaborations recently have performed measurements of 
either multiplicity (such as net charge, $K/\pi$ ratio) 
or mean transverse momentum
fluctuations~\cite{mitchelQM2004,appelsQM2004}. 
The interest was driven by a possibility to observe 
a dramatic change in fluctuation pattern with collision energy
and/or centrality indicating the phase transition. 
There exist theoretical prediction 
both for an increase of fluctuations~\cite{shuryak-stephanov} as well 
as for decrease in fluctuations~\cite{jeon-koch,asakawa} (the latter
is due to an increase in number of effective degrees of freedom
in the QGP).
At this time many new observables were introduced in the analysis.
Unfortunately, the various observables used by experiments almost
render a valid comparison of the results difficult or even impossible.
The reason for that is a persistent confusion about properties of different
observables used to measure fluctuations. 
Due to the importance of  inter experiment
comparison of the results, in the next section I try to
give a short summary of the question along with my own recommendations for 
future analyzes. This part is somewhat technical, and a
reader more interested in physical results may skip it and start
with section II.

\section{Correlation functions and fluctuations}
In a multi-particle process the correlations in particle production
are described via two and many particle densities~\cite{Whitmore,Foa}. 
In this talk I use only  single and two-particle densities;
those are normalized respectively by the mean multiplicity and the mean number 
of pairs in a given region of momentum space:
\be
\int_\DY dx \rho^{(1)}(x)=\mean{n}
;\;\;\;
\int_\DY dx_1 \int_\DY dx_2 \,\rho^{(2)}(x_1,x_2) = \mean{n(n-1)}.
\ee
Particle densities in general depend on a particle three momentum, 
e.g.  rapidity, transverse momentum, and azimuthal angle, 
some of the arguments can be integrated out; for brevity I use just 
$x$.
From $\rho^{(1)}(x)$ and $\rho^{(2)}(x_1,x_2)$
one constructs different types of correlation functions:
\be
C(x_1, x_2) \equiv \rho^{(2)}(x_1,x_2) -
\rho^{(1)}(x_1) \, \rho^{(1)}(x_2);\; 
B(x_1, x_2) \equiv \frac{C(x_1, x_2)}{\rho^{(1)}(x_1)};\;
R(x_1, x_2) \equiv \frac{C(x_1, x_2)}{\rho^{(1)}(x_1) \rho^{(1)}(x_2)}
\ee
Roughly speaking $C(x_1, x_2)$ has a meaning of a distribution of correlated
pairs.
The correlation function  $B(x_a, x_b)$ 
corresponds to the distribution of particles $b$ under condition that particle
$a$ is found at $x_a$. Often, a particle of type $a$ is called
a ``trigger'' particle and particle  of type $b$ is called
an ``associated'' particle. 
When particles $a$ and $b$ carry opposite charges, such
as electric, baryon, or strangeness,
$\int_\infty^\infty B(x_a,x_b) dx_b =1$,
as there must be one  associated  particle somewhere in the momentum space 
that ``balances'' the charge of the trigger particle.  
Then  $B(x_a, x_b)$ is often referred to as {\em balance
function}.
The correlation function $R(x_1,x_2)$
may be interpreted as the ``probability'' that a given pair is
correlated. This type of the correlation function is used most
often, as it is easiest to measure.  Defined
as a ratio, it is one of the so-called {\it robust} quantities, which do not 
depend on single particle reconstruction efficiencies. 

Studying the correlations in nuclear collisions one usually is interested
if those are different from the correlations in elementary collision.
If one superimposes a few, $N_{coll}$,
nucleon-nucleon collisions together:
\be
\rho^{(1),AA}(x)=N_{coll} \, \rho^{(1),NN}(x)
\ee
\be
\rho^{(2),AA}(x_1,x_2)=N_{coll} \, \rho^{(2),NN}(x_1,x_2)
+N_{coll}(N_{coll}-1)\,
 \rho^{(1),NN}(x_1) \rho^{(1),NN}(x_2).
\ee
Then, the balance function, $B(x_1,x_2)$, does not change at
all, and the signal in $R(x_1,x_2)$ gets diluted:
$R^{AA}(x_1,x_2) = R^{NN}(x_1,x_2)/N_{coll}$.

All of the above correlation functions can be calculated 
at fixed multiplicities. 
In this case they usually carry subscript $n$, e.q. $C_n(x_1,x_2)$,
and called {\it semi-inclusive}. 
In nuclear collision analyses 
it is not possible to fix the multiplicity of
individual nucleon-nucleon collisions, and one should compare
the results with 
{\em inclusive} $NN$ correlation function, although one should be careful
comparing very peripheral collisions selected on the basis of total
multiplicity. 
In this case the approximation of correlations in individual
$NN$ collision by inclusive correlation function can break down.

\subsection{Fluctuations}

Many different quantities have been suggested for fluctuation
analyses.
Before further discussion, 
let us formulate in general terms what constitutes a good observable. 
It should be
(i) sensitive to the physics under study, 
(ii) it should be defined
at the ``theoretical'' level, be apparatus/experiment independent,
(iii) have clear physical meaning, and 
(iv) it should not be limited
in scope, provide new venues for further study.
If most of the proposed observables do satisfy the first requirement, very
few satisfy (ii)--(iv). 
It made basically impossible to compare many of
the results, even published, one to another. 
Comparing different observables I argue that the 
``old fashion'' correlation functions, and, similar
the two particle transverse momentum correlations, $\mdptdpt$,  are the best in
many respects.

\subsubsection{Multiplicity fluctuation measures}
How the two, correlations and fluctuations, 
are related to each other?
Let us look at the so-called reduced variance:
\be
\omega_n \equiv \frac{\sigma_n^2}{\mean{n}}=1+\mean{n} \tilde{R}_\DY;\;\;
\tilde{R}_\DY=\frac{\int_\DY dx_1 \int_\DY dx_2 \rho_2(x_1,x_2)}
{\int_\DY dx_1 \int_\DY dx_2 \rho_1(x_1) \rho_1(x_2)},
\ee
where the last equation shows the relation to two particle density.
If $\rho_1(x_1)$ does not vary much over the region $\DY$,
$\tilde{R}_\DY$ gives an {\em average} of the correlation 
function $R(x_1,x_2)$ over the momentum region used in the analysis.
It follows that the reduced variance deviates from
unity (which is interpreted as the value 
for {\it statistical} fluctuations) if and
only if the correlation term $\tilde{R}$ equals zero. 

Note that $\omega$ depends on the product of $\mean{n}$ and $\tilde{R}$. 
Given $\tilde{R}$ is a robust quantity, the mean event 
multiplicity is not, it depends on detector efficiency, 
track quality cuts, etc. 
Another feature of $\omega$ is that it
can be different even if the underlying correlations are the same. 
Let us compare fluctuations of charged particles 
with fluctuations of only negative (or only
positive) particles. Even if the correlations between particles have
no any dependence on the particle charge, the reduced variance would show
different values simply because only ``half'' of the particles are
being considered.
Lastly note that the reduced variance can not be independent of
the size of the rapidity region $\DY$ as it sometimes assumed 
(it is possible only if $R(x_1,x_2)\propto \delta(x_1-x_2)$
which is not physical) as often people compare its 
values at different $\DY$.

Why one want to study  $(\mean{n}  R)$ instead of just $R$? The reason 
is that this product should stay
constant as function of centrality in a case when $AA$ collision is
considered as a superposition of independent nucleon-nucleon
collisions. 
In this case the correlation functions scales as $1/N_{coll}$,
$\mean{n} \propto N_{coll}$, and the product remains constant.
The deviation from constant is easy to observe. 
Experimentally, the centrality is often obtained by measuring
multiplicity, and it was wrongly assumed that such quantities as
reduced variance  do not depend on multiplicity in general.
It would be desirable to report both quantities separately. 
Instead of $\mean{n}$ it would be better to use $dn/dy$, a quantity 
totally corrected for efficiency and only weakly dependent on $\DY$.
Besides, if $\tilde{R}$ is known, one can check how it
scales also other quantities, such 
as number of participating nucleons,
$N_{part}$, or the number of binary collisions, or something else
that might be related to the number of sources of particle production.  
\begin{figure}[htb]
\begin{minipage}[t]{77mm}
  \includegraphics[width=0.99\textwidth, height=0.81\textwidth ]{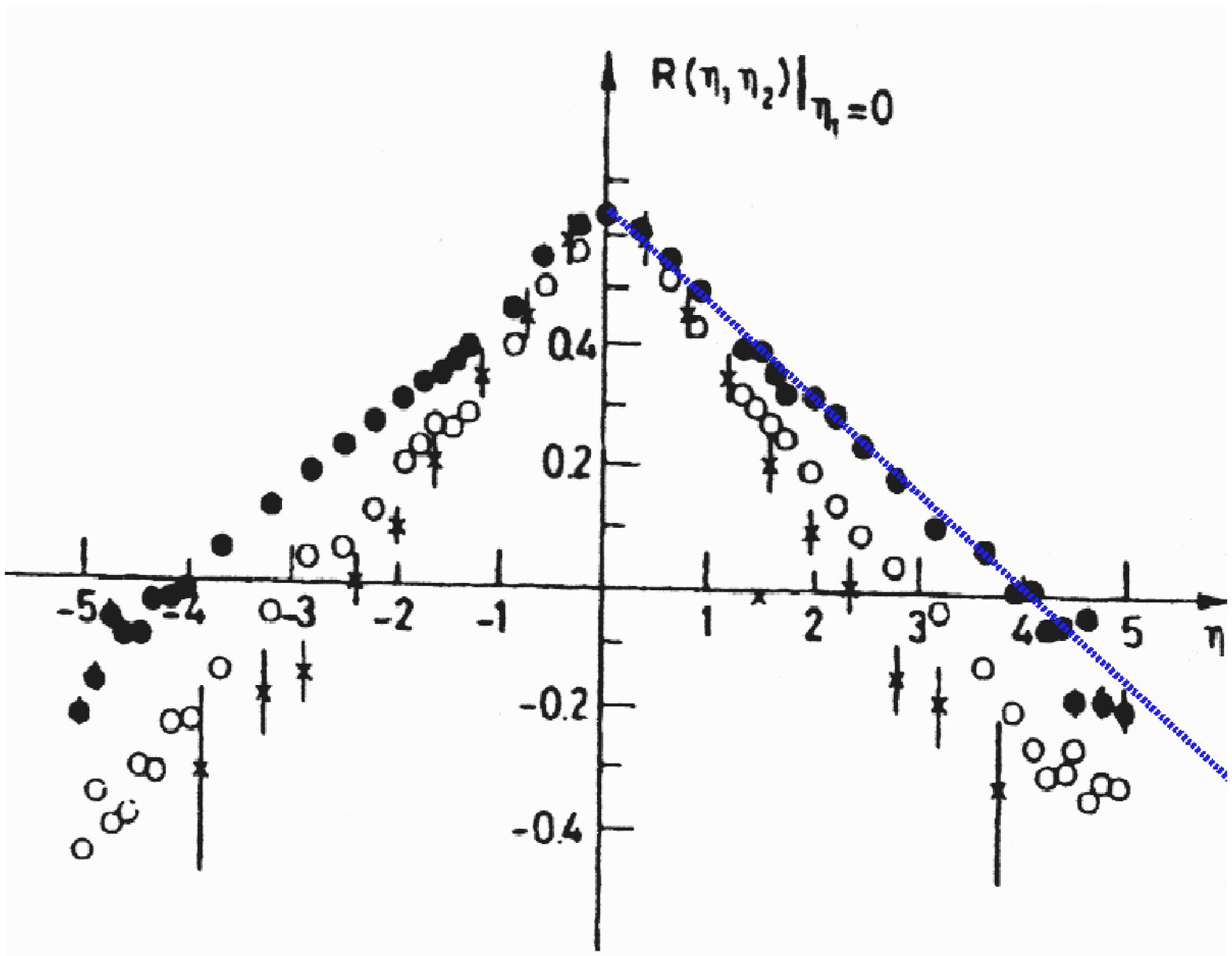}
  \caption{
The charged particle correlation function $R(\eta_1,\eta_2)$ for
  $\eta_1=0$ at various ISR energies~\cite{Foa}.
$\sqrt{s}$= 14, 23, and 63 GeV. 
} 
  \label{fISRR}
\end{minipage}
\hspace{\fill}
\begin{minipage}[t]{77mm}
 \includegraphics[width=0.97\textwidth, height=0.8\textwidth]{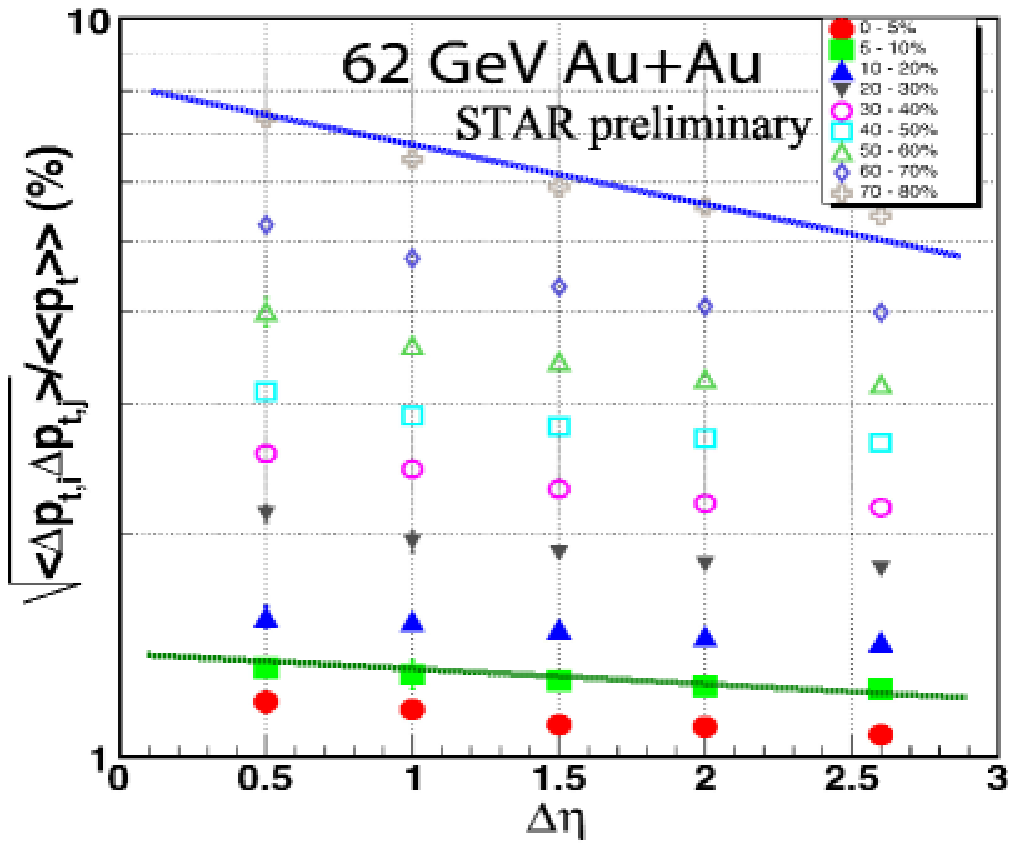}
  \caption{
$\mdptdpt$ for different centralities as function of the size of the
  pseudorapidity region~\cite{gary04}.
}
  \label{fdpt62y}
\end{minipage}
\end{figure}

To avoid the so called
``volume fluctuations'' (the effect of mixture of different impact
parameter collisions in one event sample) the fluctuations in {\it
  particle ratios} are often used, e.g. $K/\pi$, where $K$ and $\pi$
stand for the particle respective multiplicities in a given event.
Again, many ``details'' often forgotten.
First, even in the original paper~\cite{jeon-koch}, 
the fluctuations of particle ratios
are ``reduced'' (approximated) to the following
\be
\frac{\sigma^2_{K/\pi}} {\mean{K/\pi}^2} \approx
\frac{\sigma^2_{K}}{\mean{K}^2} +
\frac{\sigma^2_{\pi}} {\mean{\pi}^2}
-2\frac{\mean{K \pi}-\mean{K}\mean{\pi}} {\mean{K}\mean{\pi}}
=\frac{1}{\mean{K}}+\frac{1}{\mean{\pi}}
+(\tR_{KK}+\tR_{\pi\pi} -2 \tR_{K\pi}).
\ee
Note, that we started with statistically rather bad defined
quantity as any ratio of two random numbers is 
(what do you do if denominator is zero?) 
and ended up with a much better behaved 
quantity defined via the correlation function. 
The question is why one would want to deal with ratios when
even theoretical predictions are based on a really different 
quantity, which only approximately
corresponds to fluctuations in ratios, namely only in the case when
multiplicities of particles used in the denominator is large? 
It is much better
to report directly the correlation of the type 
$(\tR_{KK}+\tR_{\pi\pi} -2 \tR_{K\pi})$~\cite{nu-method}.
The latter is easier to measure (robust quantity) and can be 
analyzed for particles
with very low mean multiplicities (for example, $p$ and $\bar{p}$).

\subsection{Mean $p_t$ fluctuation measures}
Mean $p_t$ event by event fluctuation analyses are often performed
by calculating
the event-wise mean transverse momentum, 
an overall event average, and the variance:
\be
\mean{p_t} \equiv \frac{1}{n}\sum_i p_{t,i};\;
\mean{\mean{p_t}} \equiv \frac{1}{N_{ev}}\sum_j \mean{p_t};
\;
\sigma^2_{\mpt}=\mean{\mpt^2} -\mean{\mpt}^2 =
\frac{\sigma_{p_t,incl}^2}{n}+\frac{n-1}{n}\mean{\dptdpt},
\ee
where the first term in $\sigma^2_{\mpt}$ is defined by 
a single particle $p_t$
spectrum and is attributed to {\it statistical} fluctuations 
(the fluctuations in a case of independent particle production with 
the same single-particle distributions). 
The non-trivial fluctuations (non-statistical, sometimes called 
{\it dynamical}) are included in the second term, $\mdptdpt$~\cite{vkr}. 
All the measures of fluctuations used by different experiments 
are related to this second term (two particle transverse momentum
correlation (covariance)), although sometimes in a rather complicated
way.  In many cases such a relation involves tracking and analysis
efficiencies, and if those are not reported separately, the final 
results become not comparable to other experiments/measurements. 

If all events in the event sample have the same multiplicity, then the
equations above provide {\it semi-inclusive} mean
transverse momentum and correspondingly $\mdptdpt$. 
If the multiplicity vary within the event sample, there are two
possibilities: one can average the semi-inclusive quantities over all
events (what is used most often, I would call it event-by-event (EbyE)
average) or calculate true {\it inclusive} quantities 
\be
\meani{p_t} \equiv \frac{\sum_{events} \sum_i^n p_{t,i}}
{N_{events}\mean{n}};\;\;
\midptdpt \equiv \frac{\sum_{events} \sum_{i<j} 
(p_{t,i}-\meani{p_t}) 
(p_{t,j}-\meani{p_t}) 
}
{N_{events}\mean{n(n-1)}}.
\ee
The latter cam be written directly using two particle density 
\be
\meani{\dptdpt}=\frac{\int_\DY dx_1 \int_\DY dx_2 \rho^{(2)}(x_1, x_2)
  \dptdpt} {\mean{n(n-1)}}
\label{edpt}
\ee
Although quantitatively both definitions (EbyE and inclusive) would give almost
indistinguishable values (unless the multiplicity is too low and event
sample include events with very different multiplicities),
theoretically, it is easier to analyze inclusive correlations. Therefore
my preference would be to ``migrate'' from the EbyE
to inclusive observables.

Note also that $\mdptdpt$ can be easily generalized 
for study of correlation between particles of different
type and for differential measurements such as 
$\dpti(\eta_1,\phi_1) \dptj(\eta_2,\phi_2)$. 
Similar to the multiplicity correlation, its
centrality dependence can be tested against different hypothesis
such scaling with rapidity density, number of participants, etc.

\section{Mean $p_t$ fluctuations/correlations at RHIC}.

Recently STAR Collaboration has reported results
on integrated two particle $p_t$ correlations for Au+Au collisions at
different collisions energies~\cite{gary04}. 
Fig.~\ref{fdpt62y} shows the dependence of the two particle $p_t$
correlations on $\Delta \eta = 2|\eta_{max}|$ the size of the
pseudorapidity window taken around midrapidity.
The results for different centralities are shown starting from the most
peripheral (on the top, the strongest correlations) to the most central.
The dependence on the size of the pseudorapidity region is rather
weak. At higher collision energy ($\sqrt{s_{NN}}=200$~GeV) 
the dependence is even weaker. 
Under assumption that spectra dependence on transverse momentum and
pseudorapidity factorize, according to Eq.~\ref{edpt}, the
pseudorapidity dependence of $\mdptdpt$ should follow the one of
correlation function $R(\eta_1,eta_2)$. 
Then $\mdptdpt \propto\tilde{R}$.
For a simple estimate one can fit the correlation function by the form
$R(\Delta \eta) \propto 1-\alpha |\Delta \eta|$, which translates into
$\tilde{R} \propto 1-4/3\alpha Y$, where $Y=(\Delta \eta)_{max}/2.$
The blue lines shown in Figures 1 and 2 correspond to the same 
$\alpha = 0.16$ and describe the data well.

Note that the slope of $\mdptdpt$ dependence on $Y$ changes with
centrality. It is noticeably smaller for central collisions.
Taking into account the relation to the correlation function, it means
that the correlations in more central collision widens in $\eta$.
The same conclusion can be derived directly from the differential
measurements of the two particle $p_t$ correlation on $(\Delta \eta$,
$\Delta \phi)$ presented at this conference (see Fig.~6
in~\cite{kopytine}).
I argue below that such correlations can be due to radial transverse flow. 

\begin{figure}[htb]
\begin{minipage}[t]{81mm}
  \includegraphics[width=0.97\textwidth, height=0.8\textwidth]{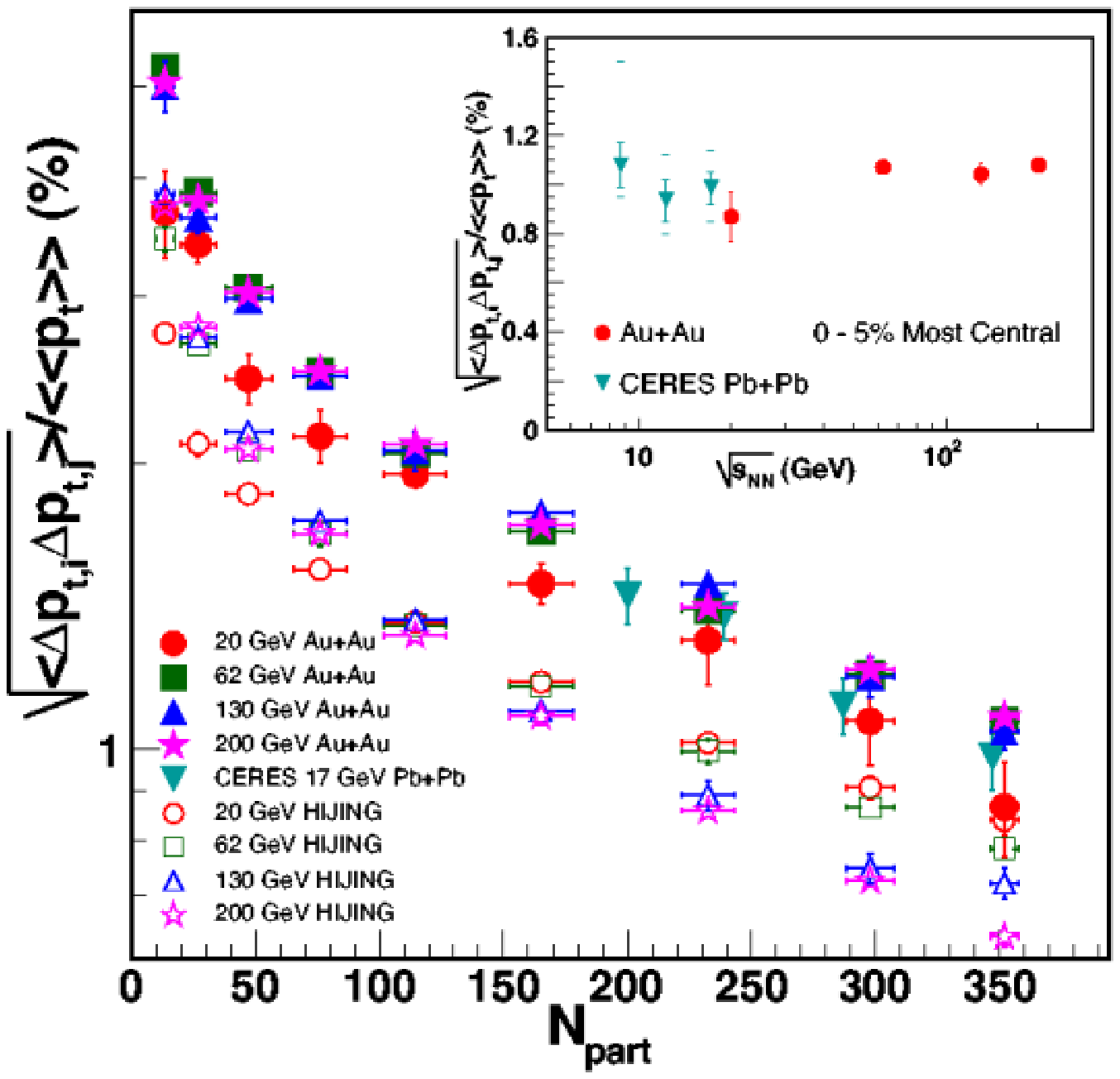}
  \caption{ 
$\sqrt{\mdptdpt} / \mpt$  as a function of centrality for
  Au+Au collisions at different collision energies~\cite{gary04}.
} 
  \label{stardptdpt}
\end{minipage}
\hspace{\fill}
\begin{minipage}[t]{78mm}
  \includegraphics[width=0.97\textwidth, height=0.84\textwidth]{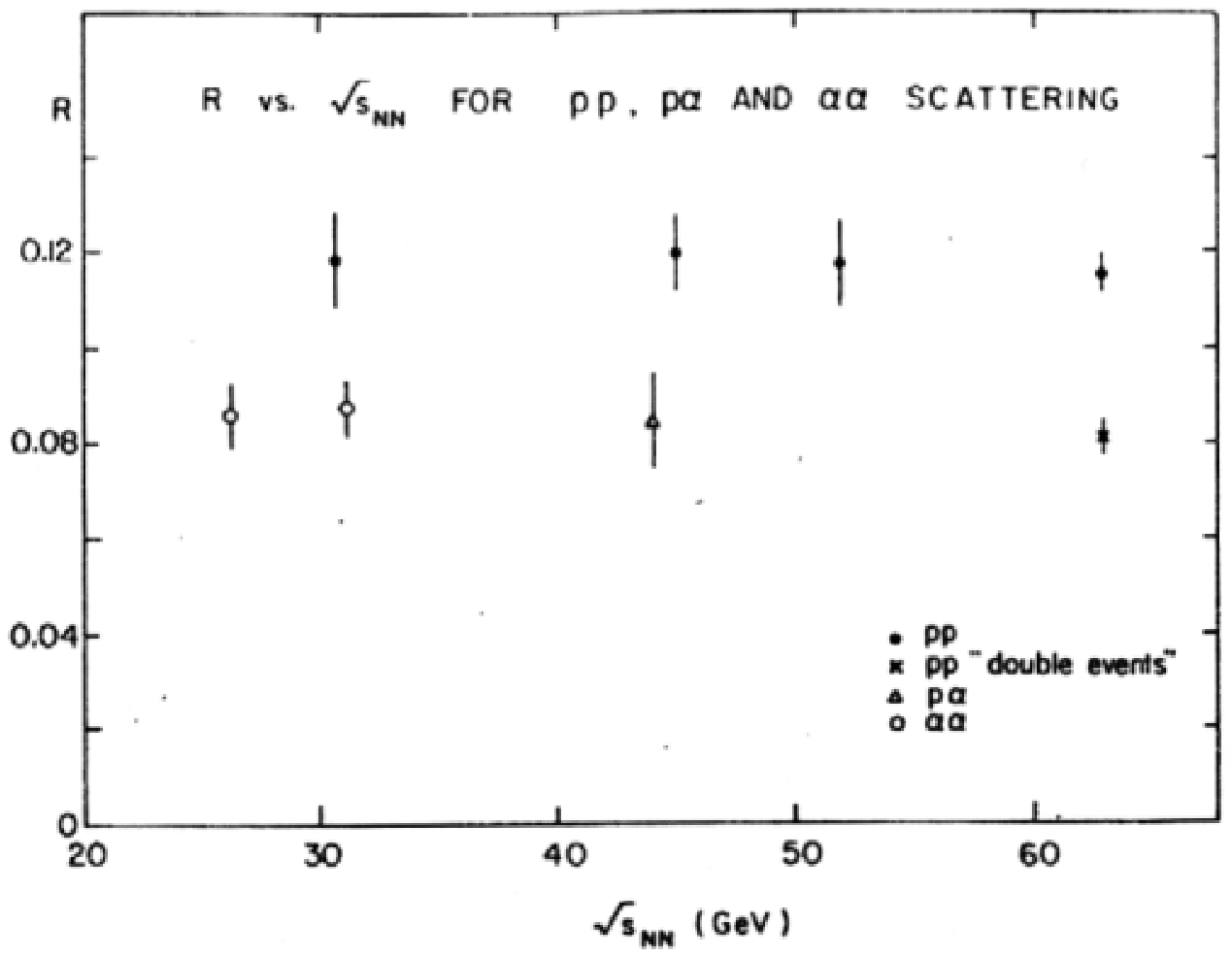}
  \caption{$\sqrt{\mdptdpt }/ \mpt $ (calculated as non-statistical 
    part in $\mpt$ event-by-event fluctuations) 
for     p+p and $\alpha+\alpha$ collisions~\cite{braune}. 
}
\label{fbraune}
\end{minipage}
\end{figure}

Fig.~\ref{stardptdpt} shows STAR results on the centrality and 
incident energy dependence of $\sqrt{\mdptdpt} /\mpt$. 
Remarkably the results in this form (two particle relative transverse 
momentum  correlations) exhibit almost no dependence on the collision energy.
The insert shows the results for the most central collision where 
  $\sqrt{\mdptdpt} /\mpt \approx 1.2$\% from lower
SPS (Pb+Pb collisions, CERES Collaboration~\cite{appelsQM2004}) 
to the top RHIC energy.    
The centrality dependence of $\mdptdpt$ roughly follows the
expectations for a dilution of the correlations as more and more
independent nucleon-nucleon collisions mixed up. 
For more detailed analysis of the centrality dependence note that
in Eq.~\ref{edpt} both numerator and denominator depend on centrality.
Taking this into account one finds
\be
\mdptdpt_{AA}  \approx
D_{N_{coll}}
\mdptdpt_{NN} 
;
\;D_{N_{coll}} =
\frac{\la n(n-1) \ra_{NN}}
{(N_{coll}-1) \la n \ra_{NN}^2 + \la n(n-1) \ra_{NN}}.
\label{edptA}
\ee 
The factor $D$ takes into account the dilution of the
correlations due to a mixture of particles from $N_{coll}$ uncorrelated 
$NN$ collisions, and
that in an individual $NN$ collision the mean number of particle pairs,
$\la n(n-1) \ra_{NN}$, on average is larger than 
$ \la n \ra_{NN}^2$. 
At ISR energies in central rapidity region
$\la n(n-1) \ra_{NN} \approx 1.66 \la n \ra_{NN}^2$~\cite{Foa} 
(see Fig.~\ref{fISRR}).   
Taking into account the dilution factor $D_{N_{coll}}$ one find that 
the correlations increase for about 50\% with centrality
relative to the expectations based on the superposition of independent $NN$
collisions (see Fig.~\ref{fdptC} below).

\section{Transverse radial flow and two-particle correlations}

At the first stage of a AA collision many individual 
nucleon-nucleon collision happen. 
Parton re-interactions lead to pressure build-up and
the system undergoes longitudinal and transverse expansion.
Transverse flow in the system creates strong position-momentum correlations 
in the transverse plane: further from 
the center axis of the system a particle is produced initially,
on average the larger push it gets from 
other particles during the system evolution. 
As all particles produced in the same $NN$ collision have initially 
the {\em same spatial} position in the transverse plane, 
they get {\em on average} the same push and thus become correlated.
This picture leads to many distinctive phenomena, most 
of which can be studied by means of two (and many-) particle
correlations~\cite{voloshinR}. 

The single particle spectra are affected by radial flow  such
that the mean transverse momentum
is  mostly sensitive to the {\em average} expansion velocity squared
$
\la p_t\ra_{AA} \approx \la p_t \ra_{NN} + \alpha \la v^2 \ra,
$
(see Fig.~\ref{fmpt} below) and to
much lesser extend to the actual velocity profile (dependence of the
expansion velocity on the radial distance from the center axis 
of the system).    
The two-particle transverse momentum correlations~\cite{vkr}, 
$\mdptdpt$, measure the {\em variance}
 in collective transverse expansion velocity, and thus are more
sensitive to the actual velocity profile.
We employ a thermal model~\cite{thermal-ssh} for further calculations. 
In this model particles are produced by
freeze-out of the thermalized matter at temperature $T$, approximated 
by a boosted Boltzmann distribution. 
Assuming boost-invariant longitudinal expansion and freeze-out 
at constant proper time, one finds
\be
\frac{dn}{d\bp_t} \sim 
\int d\rho_t d\phi_b \rho_t^{2/n-1} J(\bp_{t};T,\rho_t,\phi_b);\;\;
 J(\bp_{t};T,\rho_t,\phi_b) \equiv m_t 
K_1(\beta_t) e^{\alpha_t \cos(\phi_b-\phi)},
\ee
where $\rho_t$ is the transverse flow rapidity, 
$\phi_b$ is the boost direction,
$\alpha_t=(p_t/T)\sinh(\rho_t)$, and $\beta_t=(m_t/T)\cosh(\rho_t)$.
It also assumes a uniform matter density within a cylinder, $r<R$, and a
power law transverse rapidity flow profile $\rho_t \propto r^n$.
Two particle spectrum for particles originating from the same $NN$
collision in this picture can be written as
\be
\frac{dn_{pair}}{d\bp_{t,1} d\bp_{t,2}} \sim 
\int d\rho_t d\phi_b \rho_t^{2/n-1} 
J(\bp_{t,1};T,\rho_t,\phi_b)
J(\bp_{t,2};T,\rho_t,\phi_b) .
\ee
It additionally assumes that during the expansion time (before the
freeze-out) the particles produced originally at the same spatial position
do not diffuse far one from another compared to the system size.
The results of the numerical calculations based on the above equations
are presented in Fig.~\ref{fmpt} as function of 
$\la \rho_t^2 \ra = \la \rho_t\ra^2 (4n+4)/(2+n)^2$. 
The results are shown for two different
velocity (transverse rapidity) profiles, $n=2$, and $n=0.5$.  
\begin{figure}
  \includegraphics[width=0.5\textwidth,height=0.36\textwidth]{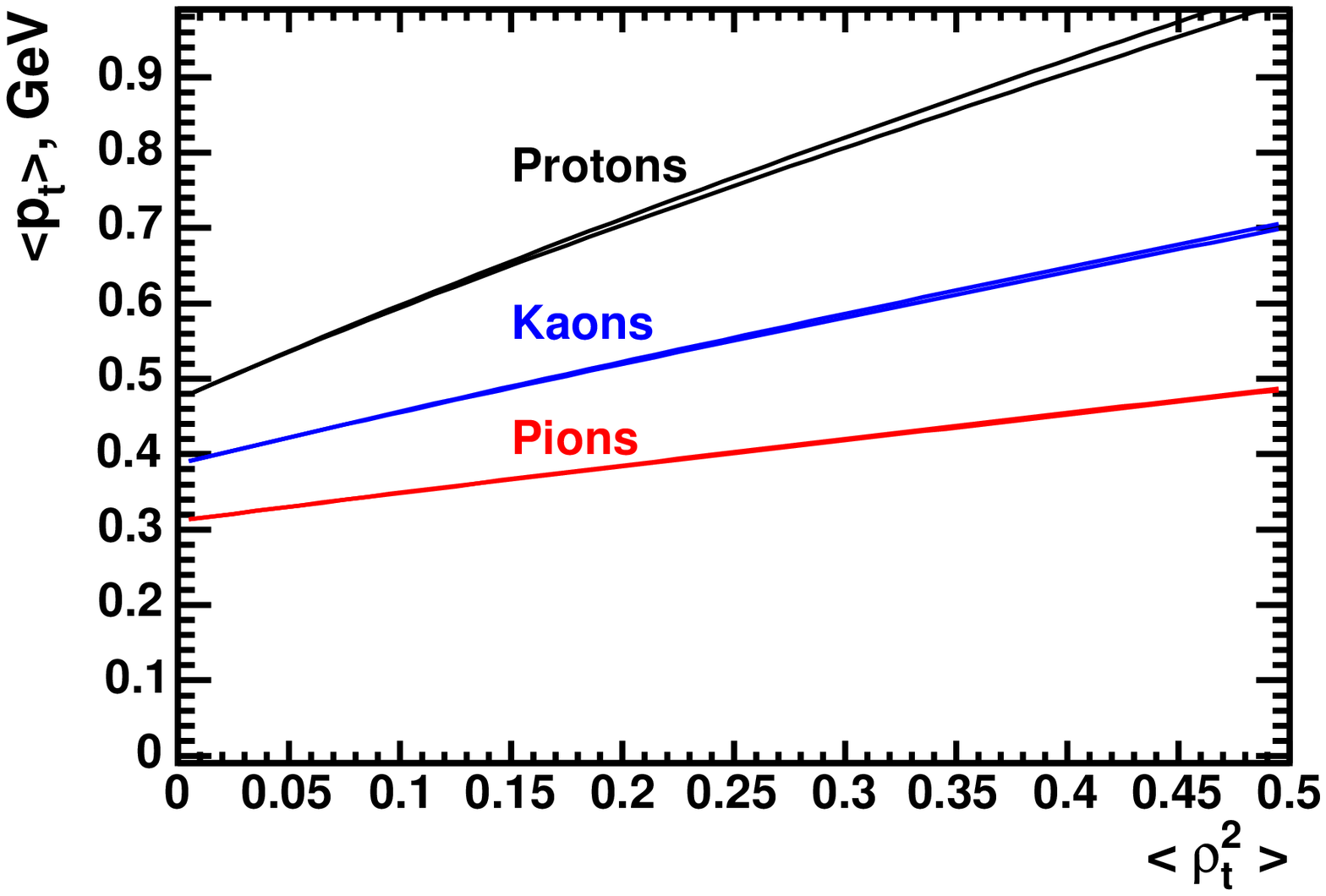}
  \includegraphics[width=0.5\textwidth,height=0.36\textwidth]{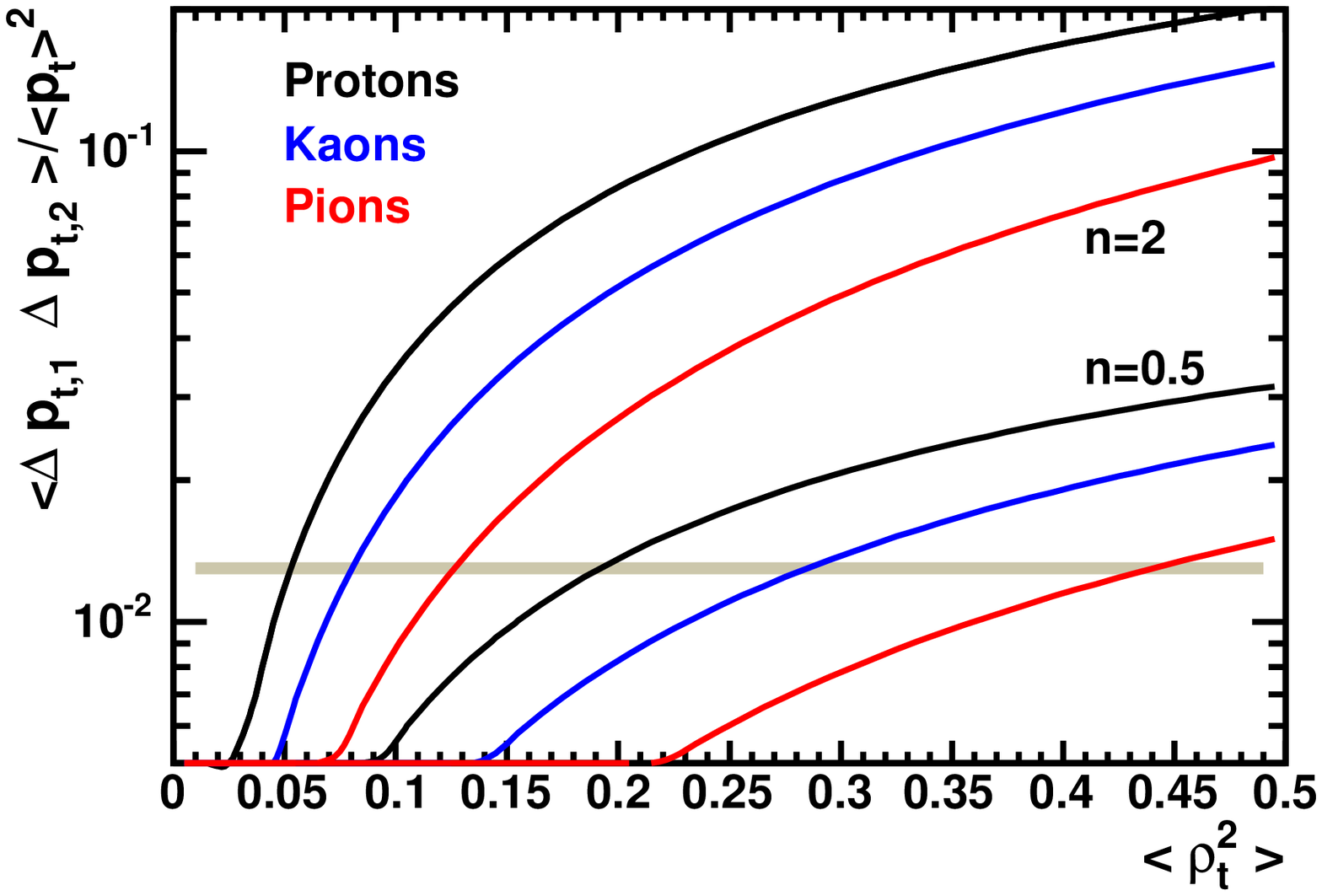}
	\vspace*{-8mm}
  \caption{
(color online) Mean transverse momentum and two particle $p_t$
  correlations according to the blast wave calculations. $T=110$~MeV.}
  \label{fmpt}
\end{figure}
One observes that indeed for all the particle types presented, $\mpt$
depends very weakly on the actual profile. On opposite, the correlations
are drastically different for two cases studied.

\begin{figure}[htb]
\begin{minipage}[t]{78mm}
  \includegraphics[width=1.0\textwidth,height=0.7\textwidth]{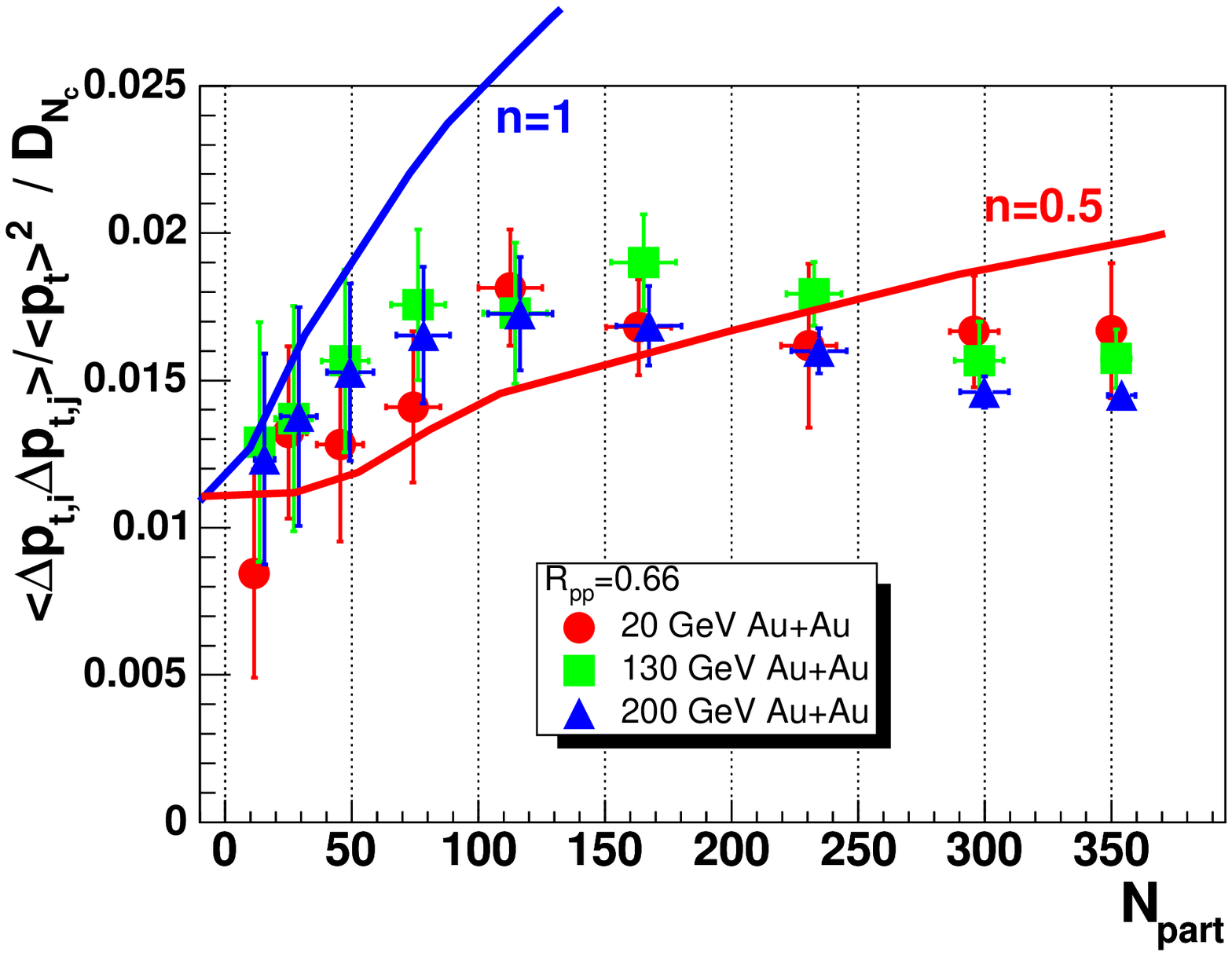}
	\vspace*{-10mm}
  \caption{
Comparison of the Blast Wave calculations for two different velocity profiles 
with preliminary STAR data~\cite{gary04}.  Relation
$\la n(n-1)\ra_{NN} =1.66 \la n\ra^2_{NN}$ has been used.
\label{fdptC}
} 
\end{minipage}
\hspace{\fill}
\begin{minipage}[t]{77mm}
  \includegraphics[width=0.95\textwidth]{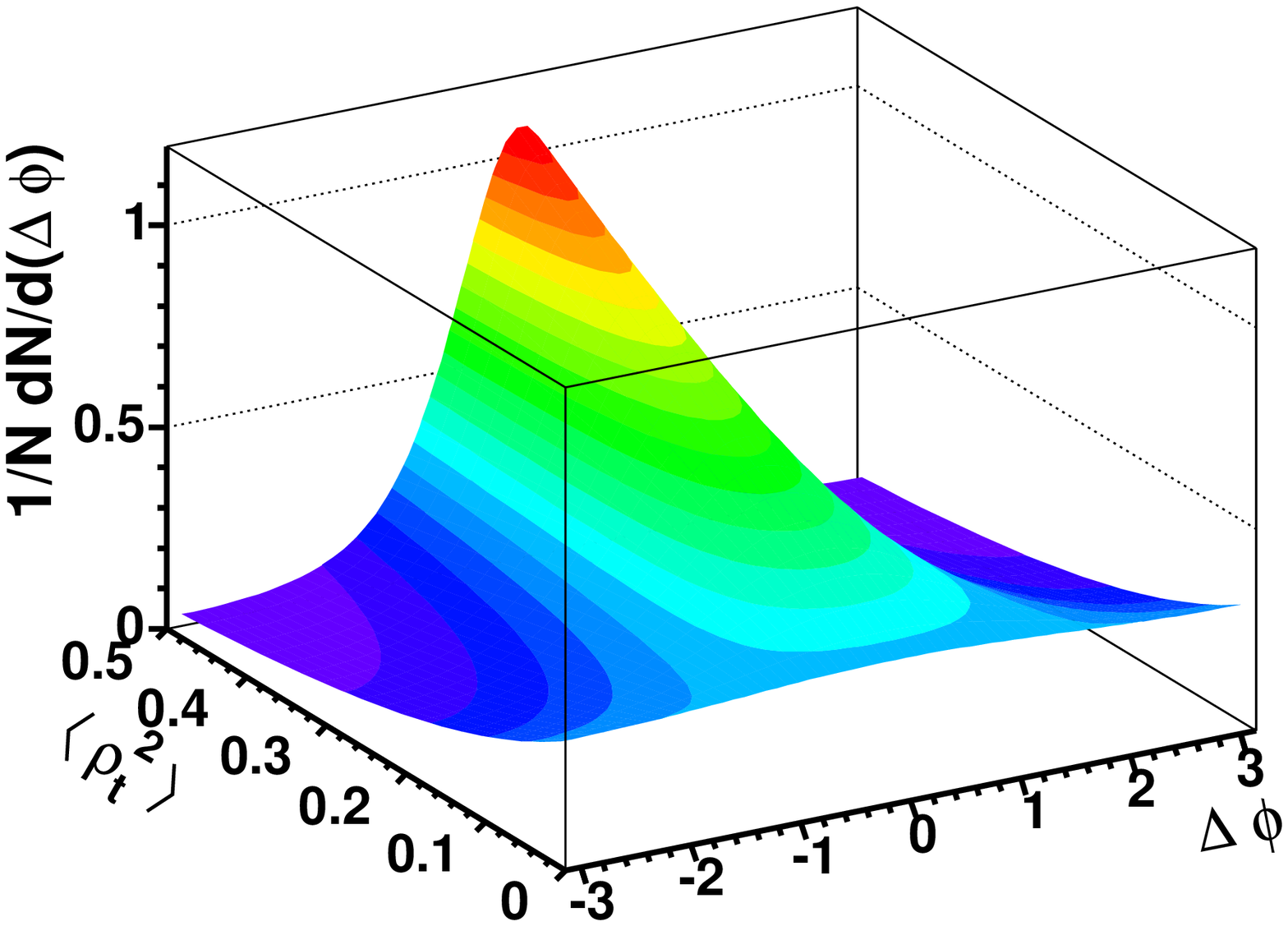}
	\vspace*{-8mm}
  \caption{(color online) Two pion $\Delta \phi$ distribution as
  function of $\la \rho_t^2 \ra$ in the Blast Wave model. 
  Linear velocity profile and
  $T=110$~MeV have been assumed.}
  \label{fdptphi}
\end{minipage}
\end{figure}

In Fig.~\ref{fdptC} we compare our estimates with 
STAR preliminary data~\cite{gary04} on
two particle $p_t$ correlations (taking into account the dilution 
factor, Eq.~\ref{edptA}). 
We use $\mean{\rho_t}$ and $T$ parameters
from~\cite{star-spectra} and assume $N_{coll}=N_{part}/2$
and $\la \dpti \dptj \ra /\la p_t\ra^2 =0.011$ (about 10\% smaller
than measured at ISR~\cite{braune} (Fig.~\ref{fbraune}). 
It is observed that the transverse flow with $n=1$ 
produces too strong correlations.

While transverse flow generates in general an elongation of the pseudorapidity 
correlations (along with narrowing in azimuthal angle)
it should lead to narrowing of the {\it charge balance 
function}~\cite{balance} due to the increase 
in mean $p_t$ as~\cite{balance2}:
$\Delta p_z=m_t \sinh(\Delta y) \approx m_t \Delta y \approx const.$
Quantitatively, the effect is consistent with experimentally observed
narrowing for about 15 -- 20\% of the balance function width 
with centrality~\cite{star-balance1} and with
centrality dependence of the net charge fluctuations~\cite{star-mult-fluc}.
As all particles from the same $NN$ collisions are pushed in the same
direction they become correlated in azimuthal space. 
The correlations can become really strong for
large transverse flow as shown in Fig.~\ref{fdptphi} 
(for particles originated from the {\em same} $NN$ collision).  
Our estimates show that the azimuthal correlations generated 
by transverse expansion could be a major contributor 
to the non-flow azimuthal correlations~\cite{star-flow-PRC}.

The above described picture of $AA$ collisions has many
interesting observable effects, only a few mentioned here. 
The picture become even richer if one looks at the 
identified particle correlations.
Many  questions require a
detailed model study, but the approach opens a potentially very interesting
possibility to address the initial conditions and the subsequent
evolution of the system created in an $AA$ collision.  

\section{Conclusion}
The correlation techniques are a very powerful tool in our quest to
understand multiparticle production in nuclear collision.
They have been proven to lead to many discoveries in the past and
promise even more to come.

{\it Acknowledgments.}
Discussions with R.~Bellwied, S.~Gavin, C.~Pruneau, S. Pratt, and
U.~Heinz are gratefully acknowledged. 
This work was supported in part by the
U.S. Department of Energy Grant No. DE-FG02-92ER40713.

 \end{document}